\documentclass[pdftex,twocolumn,epjc3,a4paper]{svjour3}

\RequirePackage[T1]{fontenc}
\smartqed  % flush right qed marks, e.g. at end of proof
\usepackage{amsmath}
\usepackage{amssymb}
\RequirePackage{mathptmx}      % use Times fonts if available on your TeX system
\RequirePackage{flushend}
\RequirePackage{mathrsfs}
\RequirePackage{dsfont}
\usepackage{units}
\usepackage{graphicx}
\usepackage[colorlinks = true,
            linkcolor = blue,
            urlcolor  = blue,
            citecolor = blue,
            anchorcolor = blue]{hyperref}
\usepackage{cite}
\usepackage{xcolor} %needed to color text 
\usepackage{gensymb}
\usepackage{upgreek}
\usepackage[switch]{lineno}
\journalname{.}

\begin{document}

\author{
  G.~Angloher\thanksref{addrMPI}\and
  G.~Benato\thanksref{addrLNGS}\and
  A.~Bento\thanksref{addrMPI,addrCoimbra}\and
  E.~Bertoldo\thanksref{t1,e1,addrMPI}\and
  A.~Bertolini\thanksref{addrMPI}\and 
  R.~Breier\thanksref{addrBratislava}\and
  C.~Bucci\thanksref{addrLNGS}\and 
  L.~Canonica\thanksref{addrMPI}\and 
  A.~D'Addabbo\thanksref{addrLNGS}\and
  S.~Di~Lorenzo\thanksref{addrLNGS}\and
  L.~Einfalt\thanksref{addrHEPHY,addrAI}\and
  A.~Erb\thanksref{addrTUM,addrWMI}\and
  F.~v.~Feilitzsch\thanksref{addrTUM}\and 
  N.~Ferreiro~Iachellini\thanksref{addrMPI,addrClu}\and
  S.~Fichtinger\thanksref{addrHEPHY}\and
  D.~Fuchs\thanksref{addrMPI}\and 
  A.~Fuss\thanksref{addrHEPHY,addrAI}\and
  A.~Garai\thanksref{addrMPI}\and 
  V.M.~Ghete\thanksref{addrHEPHY}\and
  P.~Gorla\thanksref{addrLNGS}\and
  S.~Gupta\thanksref{addrHEPHY}\and 
  D.~Hauff\thanksref{addrMPI}\and 
  M.~Ješkovsk\'y\thanksref{addrBratislava}\and
  J.~Jochum\thanksref{addrTUE}\and
  M.~Kaznacheeva\thanksref{addrTUM}\and
  A.~Kinast\thanksref{addrTUM}\and
  H.~Kluck\thanksref{addrHEPHY}\and
  H.~Kraus\thanksref{addrOxford}\and 
  A.~Langenk\"amper\thanksref{addrTUM}\and 
  M.~Mancuso\thanksref{t1,e2,addrMPI}\and
  L.~Marini\thanksref{addrLNGS,addrGSSI}\and 
  V.~Mokina\thanksref{addrHEPHY}\and
  A.~Nilima\thanksref{addrMPI}\and 
  M.~Olmi\thanksref{addrLNGS}\and
  T.~Ortmann\thanksref{addrTUM}\and
  C.~Pagliarone\thanksref{addrLNGS,addrCASS}\and
  V.~Palušov\'a\thanksref{addrBratislava}\and
  L.~Pattavina\thanksref{addrLNGS,addrTUM}\and
  F.~Petricca\thanksref{addrMPI}\and 
  W.~Potzel\thanksref{addrTUM}\and 
  P.~Povinec\thanksref{addrBratislava}\and
  F.~Pr\"obst\thanksref{addrMPI}\and
  F.~Pucci\thanksref{addrMPI}\and 
  F.~Reindl\thanksref{addrHEPHY,addrAI} \and
  J.~Rothe\thanksref{addrTUM}\and 
  K.~Sch\"affner\thanksref{addrMPI}\and 
  J.~Schieck\thanksref{addrHEPHY,addrAI}\and 
  D.~Schmiedmayer\thanksref{addrHEPHY,addrAI}\and
  S.~Sch\"onert\thanksref{addrTUM}\and 
  C.~Schwertner\thanksref{addrHEPHY,addrAI}\and
  M.~Stahlberg\thanksref{addrMPI}\and 
  L.~Stodolsky\thanksref{addrMPI}\and 
  C.~Strandhagen\thanksref{addrTUE}\and
  R.~Strauss\thanksref{addrTUM}\and
  I.~Usherov\thanksref{addrTUE}\and
  F.~Wagner\thanksref{addrHEPHY}\and 
  M.~Willers\thanksref{addrTUM}\and 
  V.~Zema\thanksref{addrMPI}
(CRESST Collaboration)
}

\institute
{Max-Planck-Institut f\"ur Physik, D-80805 M\"unchen, Germany \label{addrMPI} \and
INFN, Laboratori Nazionali del Gran Sasso, I-67100 Assergi, Italy \label{addrLNGS} \and
Comenius University, Faculty of Mathematics, Physics and Informatics, 84248 Bratislava, Slovakia \label{addrBratislava} \and
Physik-Department and ORIGINS Excellence Cluster, Technische Universit\"at M\"unchen, D-85747 Garching, Germany \label{addrTUM} \and
Institut f\"ur Hochenergiephysik der \"Osterreichischen Akademie der Wissenschaften, A-1050 Wien, Austria\label{addrHEPHY} \and
Atominstitut, Technische Universit\"at Wien, A-1020 Wien, Austria \label{addrAI} \and
Eberhard-Karls-Universit\"at T\"ubingen, D-72076 T\"ubingen, Germany \label{addrTUE} \and
Department of Physics, University of Oxford, Oxford OX1 3RH, United Kingdom \label{addrOxford} \and
also at: LIBPhys-UC, Departamento de Fisica, Universidade de Coimbra, P3004 516 Coimbra, Portugal \label{addrCoimbra} \and
also at: Walther-Mei\ss ner-Institut f\"ur Tieftemperaturforschung, D-85748 Garching, Germany \label{addrWMI} \and
also at: Excellence Cluster Origins, D-85748 Garching, Germany \label{addrClu} \and
also at: GSSI-Gran Sasso Science Institute, I-67100 L'Aquila, Italy \label{addrGSSI} \and
also at: Dipartimento di Ingegneria Civile e Meccanica, Universitá degli Studi di Cassino e del Lazio Meridionale, I-03043 Cassino, Italy\label{addrCASS}
}

\thankstext[$\star$]{t1}{Corresponding authors}
\thankstext{e1}{bertoldo@mpp.mpg.de}
\thankstext{e2}{mancuso@mpp.mpg.de}

\title{Probing spin-dependent dark matter interactions with $^6$Li}

\date{Received: date / Accepted: date}

\maketitle

\begin{abstract}
CRESST is one of the most prominent direct detection experiments for dark matter particles with sub-GeV/c$^2$ mass. One of the advantages of the CRESST experiment is the possibility to include a large variety of nuclides in the target material used to probe dark matter interactions. In this work, we discuss in particular the interactions of dark matter particles with protons and neutrons of $^{6}$Li. This is now possible thanks to new calculations on nuclear matrix elements of this specific isotope of Li. To show the potential of using this particular nuclide for probing dark matter interactions, we used the data collected previously by a CRESST prototype based on LiAlO$_2$ and operated in an above ground test-facility at Max-Planck-Institut f\"ur Physik in Munich, Germany. In particular, the inclusion of $^{6}$Li in the limit calculation drastically improves the result obtained for spin-dependent interactions with neutrons in the whole mass range. The improvement is significant, greater than two order of magnitude for dark matter masses below 1~GeV/c$^2$, compared to the limit previously published with the same data.

\keywords{Dark Matter \and Li-6 \and Spin-Dependent \and Lithium \and Crystal}

\end{abstract}

\section{Introduction}
CRESST is a direct detection dark matter experiment located in the underground laboratory of Gran Sasso (LNGS) \cite{Abdelhameed:2019hmk}. CRESST operates cryogenic calorimeters consisting of a scintillating target crystal instrumented with a Transition Edge Sensor (TES) coupled to an auxiliary cryogenic calori-meter to measure the scintillation light, called Light Detector (LD). These detectors guarantee a low energy thre-shold along with a high energy resolution for particle interactions with the target crystal. Furthermore, the employment of scintillating crystals allows to achieve an effective particle discrimination through the simultaneous detection of the light and phonon signals. As a result of these remarkable features, CRESST is one of the leading experiments for light dark matter search.\\
Light dark matter refers to dark matter particles with masses $\lesssim1$~GeV/c$^2$. There are various theoretical scenarios which could motivate particle candidates in this mass range, such as asymmetric dark matter~\cite{Kaplan:2009ag, Petraki:2013wwa, Zurek:2013wia}, scalar dark matter~\cite{Boehm:2003hm}, and hidden sector dark matter~\cite{Feng:2008ya}. These scenarios together with the absence of positive signals from experiments specialized in the detection of dark matter particles with weak-scale coupling motivate the exploration of the low-mass parameter space. \\
The CRESST-III experiment reported its first results for spin-independent dark matter interactions with nuclei of detectors employing CaWO$_4$ target crystals~\cite{Abdelhameed:2019hmk}.  In parallel to this effort, the CRESST collaboration has also investigated new suitable crystals for its detectors, with particular focus on lithium-containing crystals~\cite{Abdelhameed:2019szb,Bertoldo:2019xuf, Abdelhameed:2020qen}. The motivation for the development of these novel cryogenic detectors stems from the ideal properties of lithium for the study of the spin-dependent interactions of sub-GeV/c$^2$ dark matter with its nuclei. As an evidence of this, those prototypes achieved competitive results studying spin-dependent interactions of dark matter particles with nuclei of $^7$Li, the most abundant isotope of lithium. \\
In this work, we calculate a new limit on spin-dependent interactions of dark matter particles with $^6$Li using the data collected with these prototypes\cite{Abdelhameed:2020qen}. This analysis is now possible thanks to a recent study~\cite{Gnech:2020qtt} on the $^6$Li nuclear ground state: there, the authors report the results obtained for the computation of proton and neutron spin matrix elements which we use in this work to calculate the expected rate of dark matter interactions with $^6$Li.

\section{Dark matter interactions with lithium}

Lithium is currently the lightest element which can be employed in a CRESST-like detector. When considering elastic interactions of dark matter particles with nuclei, its light mass offers a clear advantage for the exploration of the sub-GeV/c$^2$ dark matter parameter space over heavier elements, due to the kinematics of such interactions. However, in the case of the classic spin-independent cross-section, the expected rate of interactions for dark matter particles with nuclei of lithium is significantly lower compared to heavier targets. This is due to the fact that the expected rate is proportional to A$^2$\cite{Kurylov:2003ra}, which significantly favors heavier nuclei. Overall, the employment of lithium could still offer a slight improvement over oxygen for CRESST in the study of spin-independent interactions of very low mass dark matter particles ($\lesssim$ 0.1 GeV/c$^2$), but the main advantage is the additional sensitivity to spin-dependent dark matter interactions.\\
In order to successfully probe spin-dependent interactions with a direct dark matter search experiment, the nuclides of the experimental target must hold a non-zero nuclear spin (J$_N \neq$0)~\cite{Goodman:1984dc, Ellis:1991ef, Engel:1992bf}. Furthermore, as opposed to the spin-inde-pendent standard scenario, which does not distinguish between the type of nucleons, spin-dependent interaction are usually studied for the two different cases of \textit{proton-only} and \textit{neutron-only} interactions. These \textit{proton-only} and \textit{neutron-only} interactions are proportional respectively to the square of the mean value of proton spin $\langle S_\mathrm{p}\rangle^2$ and neutron spin $\langle S_\mathrm{n}\rangle^2$, where $\langle S_\mathrm{p}\rangle$ and $\langle S_\mathrm{n}\rangle$ are the spin  matrix  elements for protons and neutrons of a given nucleus. In practical terms, it means that a suitable nuclide is usually particularly sensitive to only one of the two types of interactions, depending on its nuclear composition.\\
There is a small amount of stable nuclides, beside $^{6}$Li, composed by an odd number of protons and an odd number of neutrons which also have the property of holding a non-zero nuclear spin. These nuclides can be used to study simultaneously spin-dependent dark matter interactions with the protons and the neutrons composing their nuclei with high sensitivity. A list of such nuclides with their natural abundance is given in Table~\ref{tab:my_label}: of these, only $^{6}$Li and $^{10}$B can straightforwardly be employed in a CRESST-like detector. \\
In the case of standard CRESST-III detectors~\cite{STRAUSS2017414} made by CaWO$_4$, the isotopes which can be employed to study spin-dependent interactions are either with significantly low natural abundance ($^{17}$O and $^{43}$Ca) or large mass ($^{183}$W). Despite these limitations, in a previous work we derived a leading limit on spin-dependent interactions of dark matter particles with neutrons of $^{17}$O nuclei below 1.5~GeV/c$^2$~\cite{Abdelhameed:2019hmk}. Thus, we found a strong motivation to find more suitable target crystals for the study of spin-dependent interactions with CRESST-like detectors.\\
There has been a general lack of interest for the calculation of spin-dependent matrix elements of these isotopes listed in Table~\ref{tab:my_label}, since, for historical reasons, the community has mostly focused on nuclides composed by an odd number of nucleons~\cite{Bednyakov:2004xq}. However, it is important to stress that $^{6}$Li is arguably the best isotope which can be employed to study spin-dependent interactions of sub-GeV/c$^2$ dark matter particles with ordinary matter. In fact, $^{6}$Li is light in mass, has a not negligible natural abundance, and, mean values for both proton and neutron spin operators extremely close to the maximum value as we can see in~\cite{Gnech:2020qtt}.\\
This recent study is providing the first precise calculation of $\langle S_{p}\rangle$ and $\langle S_{n}\rangle$, the  spin  matrix  elements arising from the \textit{proton-only} and \textit{neutron-only} interactions, for an isotope of lithium. Unfortunately, up-to-date calculations for $^{7}$Li are still missing and we will refer in this work to the ones presented in~\cite{Pacheco:1989jz, Bednyakov:2004xq}.\\
A good target material compatible with cryogenic detectors is LiAlO$_2$, which was already employed in CRESST-like detectors~\cite{Abdelhameed:2020qen}. LiAlO$_2$ contains several isotopes with J$_N \neq$0: both isotopes of lithium ($^{6}$Li, $^{7}$Li), $^{27}$Al, and, $^{17}$O. The case study for $^{7}$Li was already presented in detail in~\cite{Abdelhameed:2019szb} and now we are going to present the new results obtained with the addition of $^{6}$Li to the calculation.\\
\begin{table}[]
    \centering
    \begin{tabular}{c c}
    \hline
         Nuclide & Natural Abundance (\%) \\
         \hline
       $^{2}$H & 0.01\\
       $^{6}$Li & 7.59\\  
       $^{10}$B & 19.97\\
       $^{14}$N & 99.63\\
       \hline
    \end{tabular}
    \caption{Stable nuclides with odd-numbered neutrons and protons and non-zero nuclear spin.}
    \label{tab:my_label}
\end{table}
\section{Data}
For this analysis we use the data collected in a previous work~\cite{Abdelhameed:2020qen} with the prototype detector labelled \textit{module B}. The final spectrum obtained in this work has an energy threshold for particle interactions of $E_\mathrm{T}=$(213.02$\pm$1.48) eV. The energy threshold has been calculated using the method presented in~\cite{Mancuso:2018zoh}. The energy calibration is implemented using the~5.895~keV peak from the $^{55}$Fe source and heater pulses of four different known amplitudes to interpolate the energy calibration in the whole energy region of interest, which was chosen from $E_\mathrm{T}$ to 4~keV. A total of 22.2 hours of data without any radioactive source ("background data") were collected.\\
The exposure time after analysis cuts is  17.2 hours, which corresponds to a total exposure of 2.01$\cdot10^{-3}$ kg$\cdot$d, with an exposure for $^7$Li of 1.95$\cdot10^{-4}$ kg$\cdot$d and an exposure for $^{27}$Al of 8.22$\cdot10^{-4}$~kg$\cdot$d.\\
The data are publicly available at ~\cite{Abdelhameed:2020qen} (see ancillary files on arXiv).

\section{Dark Matter results}

\begin{figure}
\centering 
\includegraphics[width=.47\textwidth]{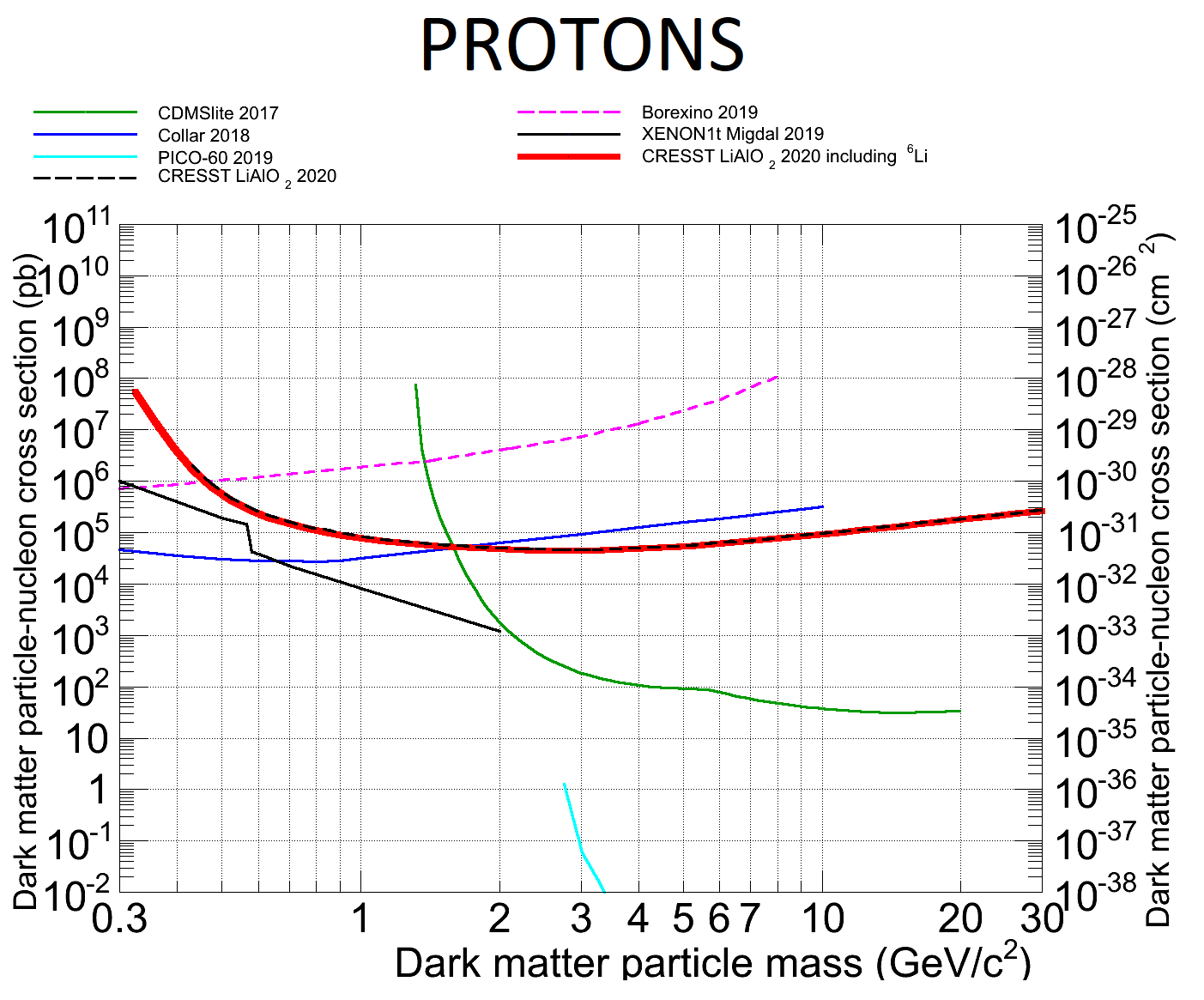}
\caption{Exclusion limits set by various direct detection experiments for spin-dependent interactions of dark matter particles with protons.The result obtained from \textit{module B} data with $^7$Li+$^{27}$Al is shown with a dashed black line~\cite{Abdelhameed:2020qen}. The new result obtained with $^6$Li+$^7$Li+$^{27}$Al is shown with a solid red line. Additionally, limits from other experiments are also shown: CDMSlite with $^{73}$Ge~\cite{Agnese17}; PICO with $^{19}$F~\cite{Amole:2019}; XENON1T (Migdal effect) with $^{129}$Xe+$^{131}$Xe~\cite{Aprilemigdal}; J. I. Collar with $^{1}$H~\cite{Collar2018}. Finally, a constraint from Borexino data derived in~\cite{Bringmann2018} is shown in black.}
\label{fig:DMLimits_p}
\end{figure}

\begin{figure}
\centering 
\includegraphics[width=.47\textwidth]{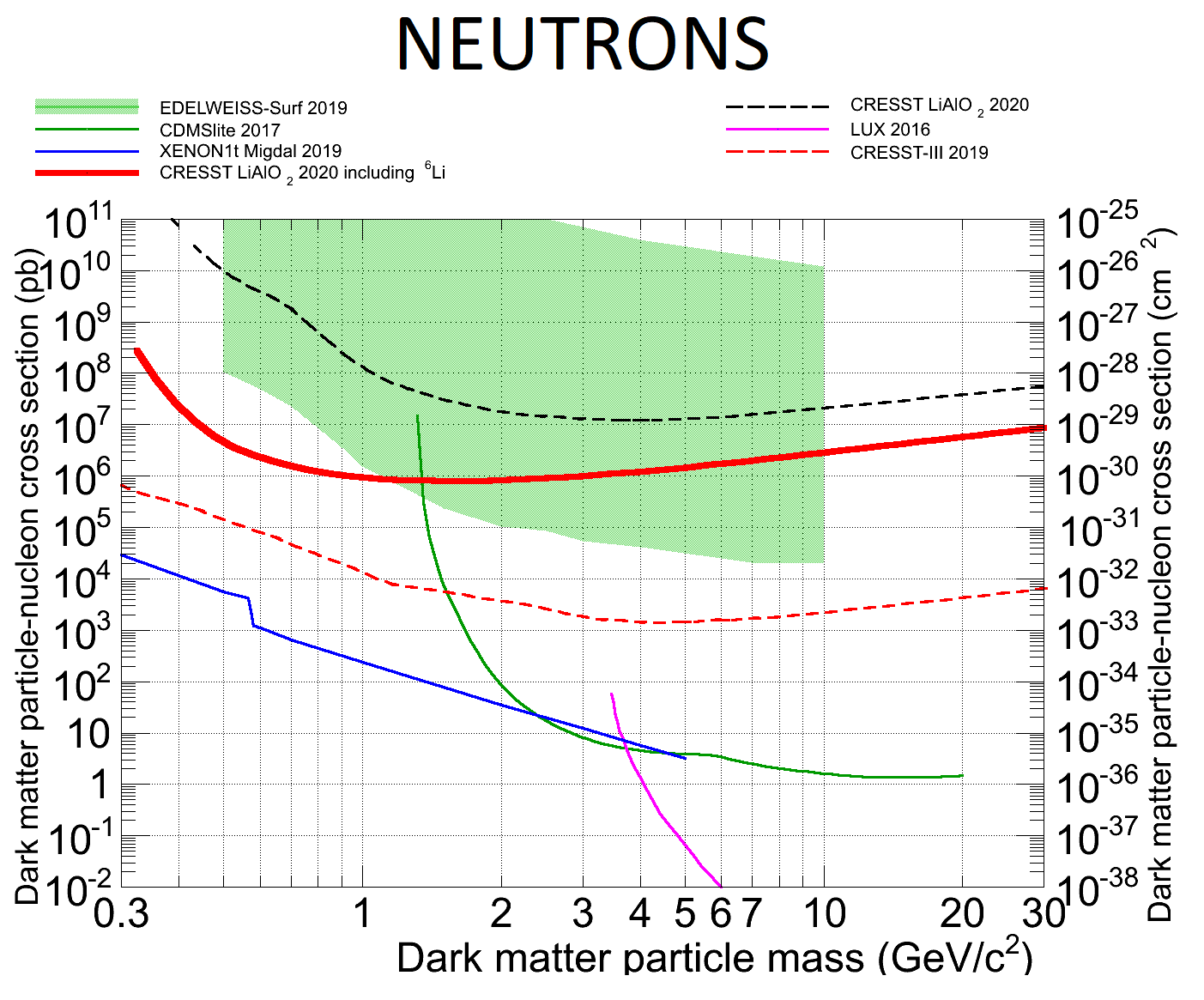}
\caption{Exclusion limits set by various direct detection experiments
for spin-dependent interactions of dark matter particles with neutrons. The result obtained from \textit{module B} data with $^7$Li+$^{27}$Al is shown with a dashed black line~\cite{Abdelhameed:2020qen}. The new result obtained with $^6$Li+$^7$Li+$^{27}$Al is shown with a solid red line. The result obtained with $^{17}$O in CRESST-III is shown with a dashed red line~\cite{Abdelhameed:2019hmk}. For comparison, limits from other experiments are also shown: EDELWEISS~\cite{Armengaud2019} and CDMSlite~\cite{Agnese17} using $^{73}$Ge, LUX~\cite{Akerib2017} and XENON1T (Migdal effect)~\cite{Aprilemigdal} using $^{129}$Xe+$^{131}$Xe.}
\label{fig:}
\label{fig:DMLimit_n}
\end{figure}

%The dark matter results presented in this work are derived by the data collected with the prototype labeled \textit{module B} in~\cite{Abdelhameed:2020qen}. To summarize, 22.2 hours of data were collected, in an above ground laboratory employing a 2.8~g LiAlO$_2$ crystal instrumented with a TES. The exposure time after analysis cuts is  17.2 hours, which correspond to a total exposure of 2.01$\cdot10^{-3}$ kg$\cdot$day, with an exposure for $^7$Li of 1.95$\cdot10^{-4}$ kg$\cdot$day and an exposure for $^{27}$Al of 8.22$\cdot10^{-4}$~kg$\cdot$day. \\
To be fully consistent, the new results are calculated in the same fashion as the one presented in the previous work, with the only difference being the addition of the cross section for dark matter interactions with $^{6}$Li alongside to the ones for $^{7}$Li and $^{27}$Al. As such, the exposure of  $^{6}$Li is equal to 1.60$\cdot10^{-5}$~kg$\cdot$d.\\
The limits are calculated using Yellin's optimal interval me-thod \cite{Yellin02, Yellin08} and are shown in Figures~\ref{fig:DMLimits_p} and \ref{fig:DMLimit_n}. The calculation adopts the standard dark matter halo model, with a Maxwellian velocity distribution and a local dark matter density of $\rho_\text{DM} = \unit[0.3]{(GeV/ \- c^{2}) /cm^{3}}$~\cite{Salucci2010}, a galactic escape velocity of $v_\text{esc} = \unit[544]{km/s}$~\cite{Smith2006} and $v_\odot = \unit[220]{km/s}$ for the solar orbit velocity~\cite{Kerr1986}. To calculate the \textit{neutr\-on-only} and \textit{proton-only} exclusion limits for $^6$Li, we adopted $\langle S_\mathrm{n}\rangle = \langle S_\mathrm{p}\rangle = 0.472$~\cite{Gnech:2020qtt}.\\
As is possible to see, the addition of $^{6}$Li slightly improves the result obtained for \textit{proton-only} interactions, due to a modest increase in the overall exposure. Instead, for \textit{neutron-only} interactions we can observe a clear improvement (Figure~\ref{fig:DMLimit_n}), by at least one order of magnitude over the whole mass range with respect to the previous result~\cite{Abdelhameed:2020qen}. In particular, there is a drastic improvement in the sub-GeV/c$^2$ dark matter mass region, by more than 2 orders of magnitude at 1.0~GeV/c$^2$ and over 3 orders at 0.4~GeV/c$^2$. This is exclusively due to the added sensitivity provided by the inclusion of $^{6}$Li in the limit calculation. \\
For a dark matter particle mass of 1.0~GeV/c$^2$ the limit on the cross section reaches 7.59$\cdot$10$^{-32}$~cm$^2$ for \textit{proton-only} interactions and 9.23$\cdot$10$^{-31}$~cm$^2$ for \textit{neutron-only} interactions.

\section{Conclusions}
In this work we present a detailed analysis of the potential to probe spin-dependent dark matter interactions with $^6$Li for both \textit{proton-only} and \textit{neutron-only} interactions. Since $^{6}$Li is always present in nature alongside $^{7}$Li, an ideal candidate to probe \textit{proton-only} spin-dependent interactions, the major improvement offered by the presence of $^{6}$Li is a clear enhancement in sensitivity for \textit{neutron-only} interactions.\\
As such, we have shown a real case scenario with a CRESST-like detector employing a LiAlO$_2$ target crystal: the total improvement with the simple addition of $^{6}$Li in the limit calculation is equal to more than two orders of magnitude for a dark matter particle mass of 1.0~GeV/c$^2$ compared to the result previously obtained.\\
This result strengthens the case for lithium-containing crystals as the ideal target material for cryogenic detectors designed to probe sub-GeV/c$^2$ spin-dependent dark matter interactions with ordinary matter.

\begin{acknowledgements}
We thank Alex Gnech for the useful discussion which gave a fundamental contribution to this work.\\ This work is supported through the DFG by SFB1258 and the Origins Cluster, and by the BMBF05A17WO4.

\end{acknowledgements}

\bibliographystyle{h-physrev}
\bibliography{biblio.bib}

\end{document}